# Cyber-attack Mitigation and Impact Analysis for Low-power IoT Devices

Ashutosh Bandekar, Ahmad Y Javaid*, *Member IEEE*

*Abstract*— **Recent years have seen exponential development in wireless sensor devices and their rebirth as IoT, as well as increased popularity in wireless home devices such as bulbs, fans, and microwave. As they can be used in various fields such as medical devices, environmental studies, fire department or military application, etc., security of these low powered devices will always be a concern for all the user and security experts. Users nowadays want to control all these "smart" wireless home devices through their smartphones using an internet connection. Attacks such as distributed attacks on these devices will render the whole system vulnerable as these attacks can record and extract confidential information as well as increase resource (energy) consumption of the entire network. In this paper, we propose a cyber-attack detection algorithm and present an impact analysis of easy-to-launch cyber-attacks on a low-power mote (Z1 Zolertia) as a model IoT device. We also present detailed results of power consumption analysis with and without attack along with when the mitigation algorithm for intrusion detection is implemented.**

***Keywords— IoT network security, Cyber-attack mitigation, intrusion detection system***

## I. Introduction

Internet of Things (IoT) devices provide users several advantages including being easy to use, compact and wireless in connectivity. These devices comprise of sensors such as temperature, humidity, and sound, and are usually part of a network allowing it to be controlled through a centralized device such as a smartphone or a PC. Zolertia Z1, Sky Mote, and T-mote are a few examples of wireless sensor network devices of the first generation. The newer generation of these devices, most commonly known as IoT devices, are packaged as easy to use devices for use as smart homes devices. Also, companies such Amazon has dash buttons which are equipped with these IoT devices [1]. As these devices are compact and provide user to control through easily accessible mobile devices, they have a wide variety of applications such as health monitoring systems, hands-free household device control, and military applications.

However, according to some studies, 70 percent of these devices have many vulnerabilities [2]. We are using Zolertia Z1 in our experiment as this device works on the same principle as the newer generation of IoT devices works. As these devices are low-powered and have limited computational capabilities, the impact of attacks such as wormhole or flooding will have a disastrous effect on the entire network. To prevent these attacks, there is a need of a power efficient intrusion detection system (IDS). Such an IDS can monitor the whole network constantly for malicious activity or violation as a result of an intrusion. IDS can either be an additional hardware or an algorithm. If IDS is in the form of hardware, then there is a risk of increased power consumption within the IoT network. Recently, there has been numerous cases of huge DDoS attacks using millions of IoT devices [3]. In the specifically cited case, millions of IoT devices, packaged with an insecure operating system, were used to send requests to a DNS provider called Dyn, leading to disruption in services of providers like Netflix, CNN, and Twitter [4].

Therefore, it can be concluded that manufacturers have not been paying much attention to the security of these IoT devices while focusing on fast-track development, marketing and delivery to the market. This also necessitates the development of a secure communication algorithm for IoT devices which can be directly used on any of these IoT devices irrespective of their manufacturer to protect organizations as well as individuals using them for their smart homes.

### *A. Contributions*

Considering these limitations and requirements, there is a need for an algorithm that will provide security and overcome resource limitations. Such a secure message passing technique should be designed for these low powered devices along with cyber-attack detection. Clearly, implementing various secure cryptographic protocols on these devices will increase the consumption of resources on these IoT devices. This paper proposes a novel algorithm that is employed within Zolertia Z1 motes to test the efficiency of the algorithm in terms of enhancing security and reducing power consumption, as these devices work on the same principle as the newer generation of IoT devices. After result validation, we can implement the same algorithm on other popular IoT devices.

### *B. Paper outline*

The next section discusses related work in the domain and includes some of the existing implementations. In section III, the information about the technology used such as hardware, software, and environment selected for the experimental setup are discussed. Most importantly brief information about the implemented algorithm. Section IV presents various simulation and experimental results along with their comparative analysis. Finally, we conclude the paper in section V.

## II. Related Work

In this paper, we are discussing and proposing a novel cyber-attack mitigation technique for IoT device networks. The analysis conducted considers the impact of cyber-attacks and deployment of the IDS algorithm on the network comprising of IoT devices. In a previous work, a comparative analysis of energy consumption and battery life estimation for similar IoT devices was presented [5]. Similar experiment pattern is being followed in this paper to validate our algorithm. However, in this paper, the efficiency of the proposed security algorithm is discussed, and a comparison has been presented of the results after conducting the same experiment in enclosed lab space with and without IDS with an ongoing cyber attack. Most works in the area of energy consumption analysis with ongoing attacks and implemented IDS use simulation rather than real-world motes. However, a study deals with real world experiments with a Z1 and Open motes [6]. On the contrary, this work does not target security or analyzes the behavior of motes when a security algorithm is implemented. There is one security based work, but it is also limited in its analysis due to its nature of being simulation based [7].

As w are considering IoT devices constrained energy resources will always be an important aspect. There is an investigative study regarding performance measurements [8], energy consumption and computation point of view with the help of MICAz and TelosB sensors after the implementation of Elliptical curve cryptography One more similar study has been conducted considering all Mica wireless sensor network devices and TelosB sensors [9] but these devices have more computing features compared to Zolertia Z1 as it is very low powered devices consisting very limited. Also in one of the studies, there is a proposed lightweight encryption for IoT devices. [10] As these devices have many security issues if they are being used in healthcare application systems such as Wireless body area network. [11] If the power consumption of these IoT devices is considered, then there are many studies but out of them a survey of the wireless sensor network power consumption is present, but it has been conducted in a simulation environment. [12]

However, in this paper, the efficiency of the proposed security algorithm is discussed, and a comparison has been presented of the results after conducting the same experiment in enclosed lab space with and without IDS with an ongoing cyber attack. Most works in the area of energy consumption analysis with ongoing attacks. In the next section, there will be a detailed description of our technological approach towards conducting the experiments.

## III. Technical Approach

As we are conducting an impact analysis of cyber-attack and cyber-attack mitigation for low-power IoT devices, we are using Zolertia Z1 motes as our testing platform.

### A. Hardware

These devices use the second-generation MSP430, a low powered 16-bit MCU operating at 16 MHz. It has the micro-USB connector for power and debugging. It is also equipped with the CC2420 transceiver that operates at 2.4GHz and is compliant with IEEE 802.15.4, 6LowPAN, ZigBee protocols. It also comprises of other sensors such as 3-Axis gyroscope, ±2/4/8/16 g digital accelerometer (ADXL345), and digital temperature sensor (TMP102) with ± 0.5ºC accuracy [13]. These devices work in the range of 0.3-3.6V, and it can even operate on two 1.5V AA batteries [5] and code size as these IoT devices have limited memory. This OS requires kernel, libraries, the program loader and a set of processes and supports C programming language. [13]

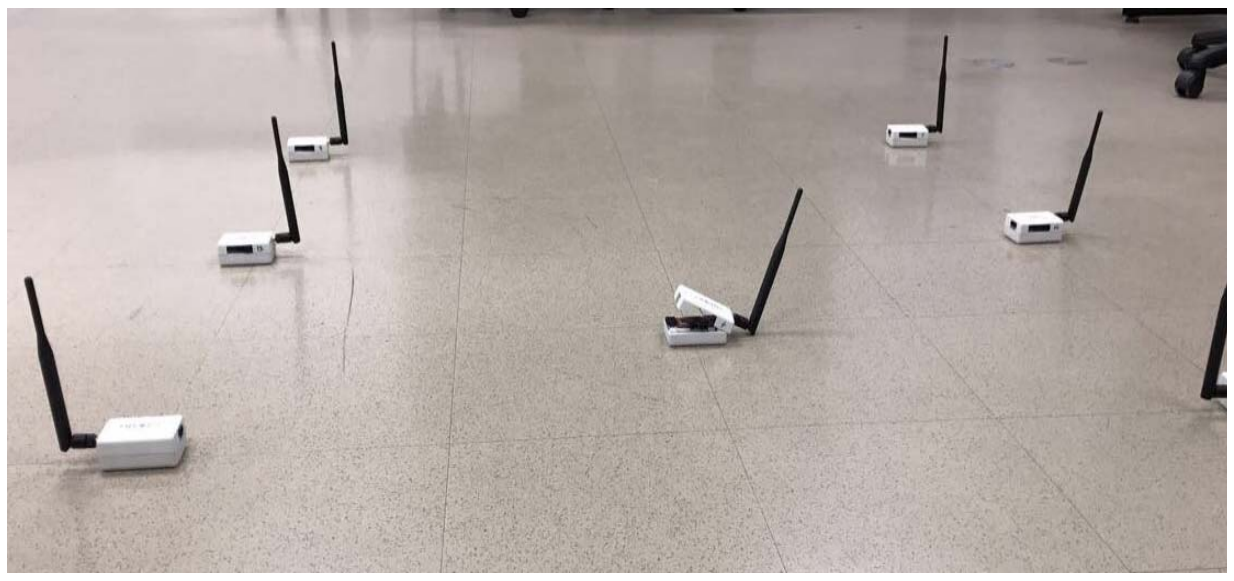

Figure 1. Topology used for the Z1 motes in real-world experiments

### B. Environment

We are analyzing the impact of cyber-attack and proposed IDS by changing different operating conditions such as basketball court, auditorium, open parking lot, and working lab space as well as in various lighting conditions. During simulation, we followed the same topology, i.e., eight motes in a grid of 15ft x 5ft and repeated the same procedure except changing the operating environment. Cooja simulation platform has limited support in terms of simulation in a different environment [14].

### C. Experimental Setup

For experimental setup, we created testbed of nine motes and out of nine, we used eight motes to create a grid of 15ft x 5ft on which we ran the broadcast and one to one communication programs. The ninth mote collected the power consumption data using the powertrace code in different operating and lighting conditions [5] [15]. This process was followed to observe the behavior of network when there is no attack. Then we echoed the same procedure by converting one of the motes from the grid to the attacker and again the congruent methodology by implementing our proposed IDS in one of the eight motes. From figure 1, we can observe the topology used for real world mote experiments. To validate the results from the real-world mote experiments, we also analyze the impact on simulation using Cooja.

### D. Software

As discussed in the previous section we considered Contiki OS for the provision of Zolertia Z1 drivers. Contiki OS is very efficient OS for low-powered devices such as Z1 mote, Sky mote, etc. For running Contiki OS require kernel, libraries, the program loader and the set of processes [15]. This OS supports C programming language, and it can run platforms such as MSP430 [14] [16]. Contiki OS is capable of implementing any topology according.

From figure 2, we can compare the protocols followed by IoT and regular IP network. In the figure, we can notice that after the presentation of the layer representation of the regular network stack is there and after that protocols followed by IoT devices and implementation of the Contiki layer.

| | ***Traditional Network*** | ***IoT Network*** |
|---|---|---|
| Application Layer | HTTP, FTP, SMTP,etc. | CoAP, MQTT, XMPP, AMQP |
| Transport Layer | TCP/UDP | UDP, DTLS |
| Network Layer | IPv6, IPv4, IPSec | IPv6/IP, 6LoWPAN |
| Link Layer | 802.3, 802.11, 802.15.4 | 802.15.4 MAC/PHY |
| Physical Layer | Ethernet, DSL, ISDN, WLAN, etc. | CC2420 |

Figure 2. Implementation in Regular Network, IoT network and Contiki layer [7] [17] [18]

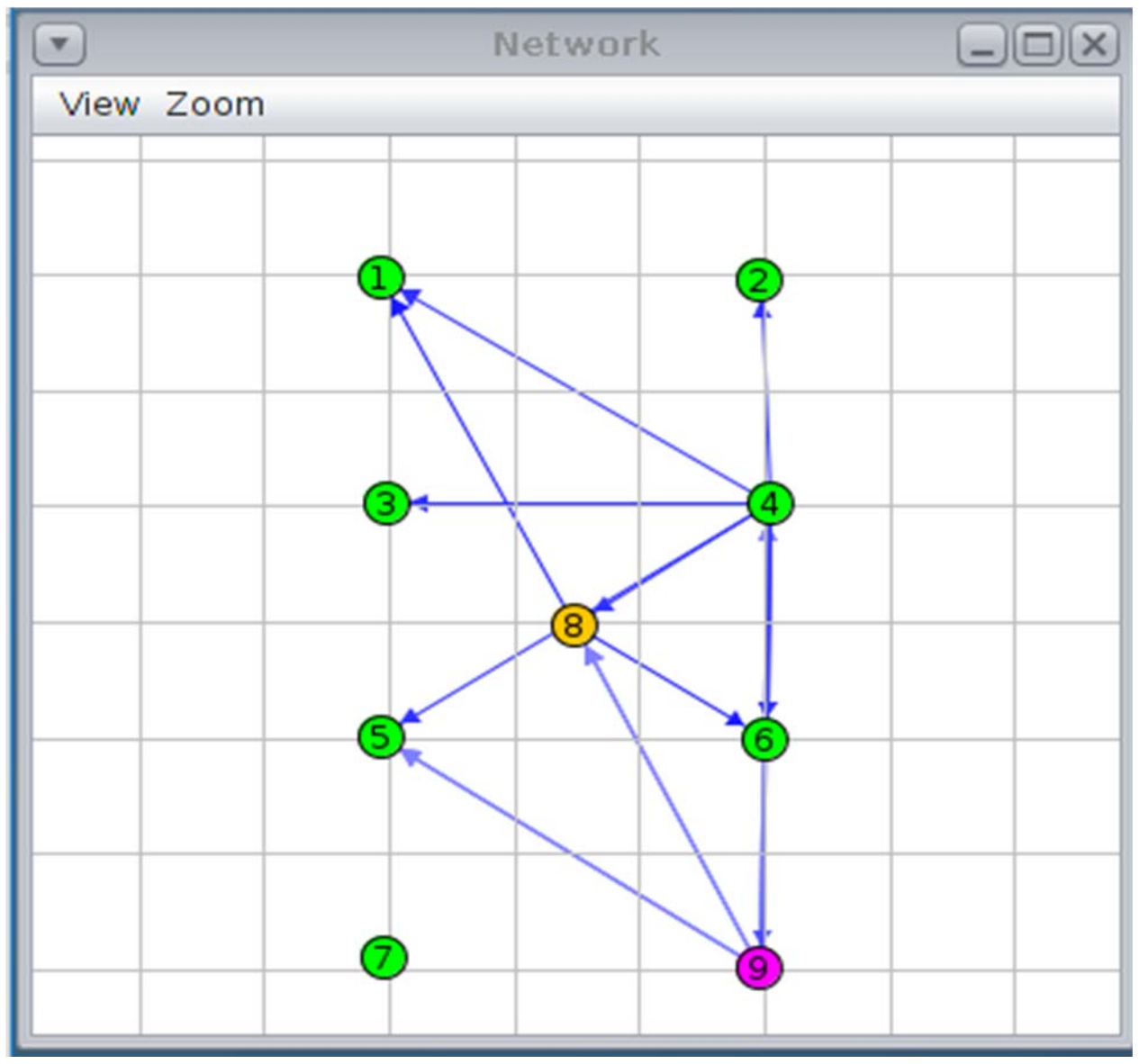

Fig 3. Topology used in simulation

We followed the same topology as we can observe from the figure 3, with which we ran simulation without any proposed implementation with the help of nine motes. We can observe the topology used in simulation in the figure. On eight out of nine motes we ran Broadcast and Unicast and on remaining one powertrace example. Then in simulation, also we employed our proposed implementation following the same pattern.

### *E. Attack Mitigation Algorithm*

To mitigate cyber-attacks, an IDS that will monitor the IoT network and detect attacks is required. To mitigate the attack and secure the network from intrusions, the proposed technique combines physical layer parameter based key generation for cryptography with packet encapsulation. The algorithm detects several types of intrusions as per our experiments with wormhole and Sybil attacks.

**Algorithm 1** Cyber-attack Mitigation in IoT Devices

**Input:** Node = N to n, Distance = di, Range = D, Vn = Victim Node

**Output:** Vn, An = attacker node

```
1:  for i = 1 to n do
2:      for j = 1 to n except i do
3:          Calculate dij = distance (Ni, Nj)
4:          D = distance through RSSI monitoring;
5:          if D>di then
6:              Attacker Flag=1
7:              set Vn = Ni
8:              Go to step 1
9:          else
10:             Attacker Flag = 0;
11:         continue
12:         end if
13:     end for
14: end for
```

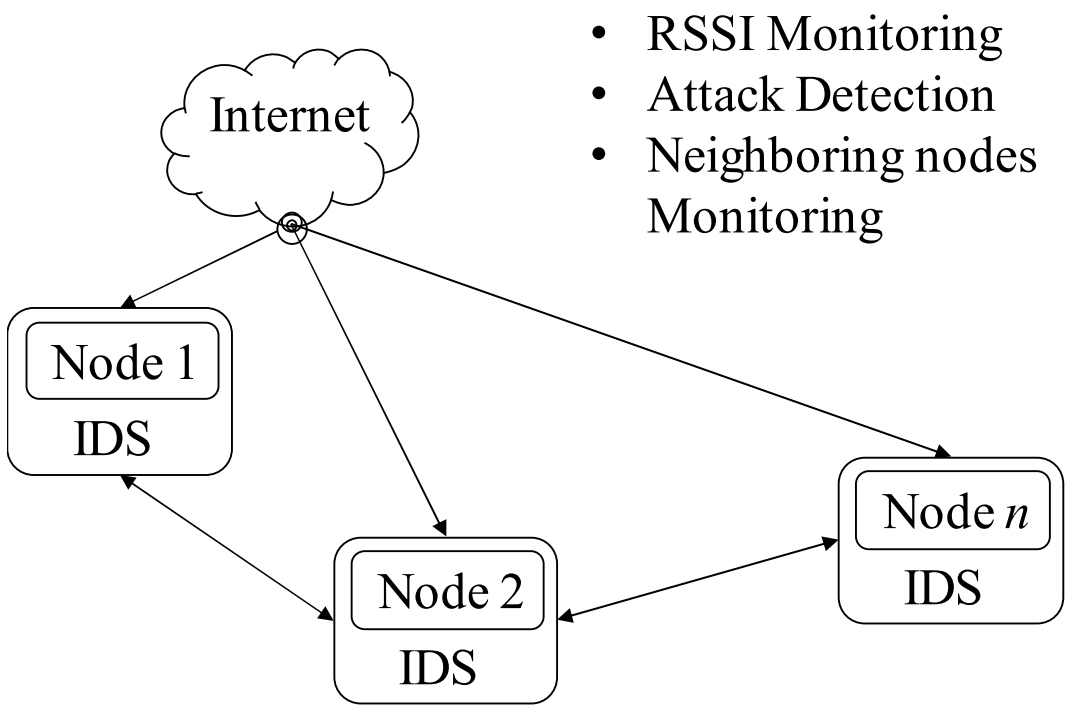

Figure 4. Architecture of the Attack mitigation algorithm

The physical layer parameters used for key generation include RSSI, location, and other channel parameters. In the figure we can observe the architecture we are proposing. We can notice that we are connecting all the nodes to the internet with the help of IPv6 protocol and during the implementation, we are considering all the key generation, neighboring nodes monitoring, RSSI collection and monitoring and attack detection.

Figure 4 we can notice the general architecture of the implemented intrusion detection system. It has one anchor node to collect, determine the neighboring node information. This architecture follows the following steps:

a. Determining the distance between each node in the network
b. Collecting the information regarding the neighboring node(s) such as the distance of the neighboring nodes from the anchor nodes.
c. If the neighboring in not in the range of the anchor node then
   i. Generate alert

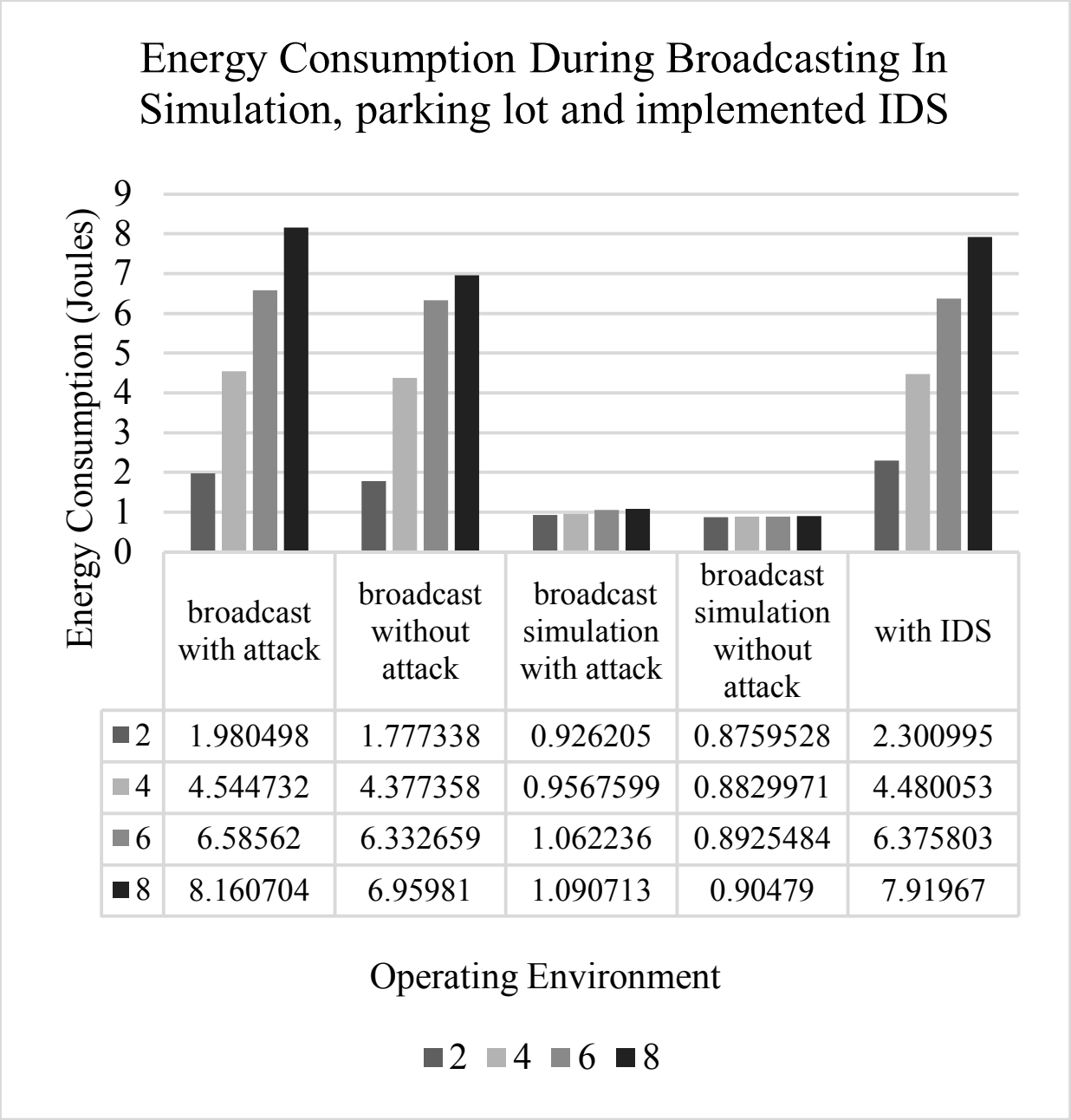

Figure 5. Energy Consumption during broadcasting in Simulation and working lab space

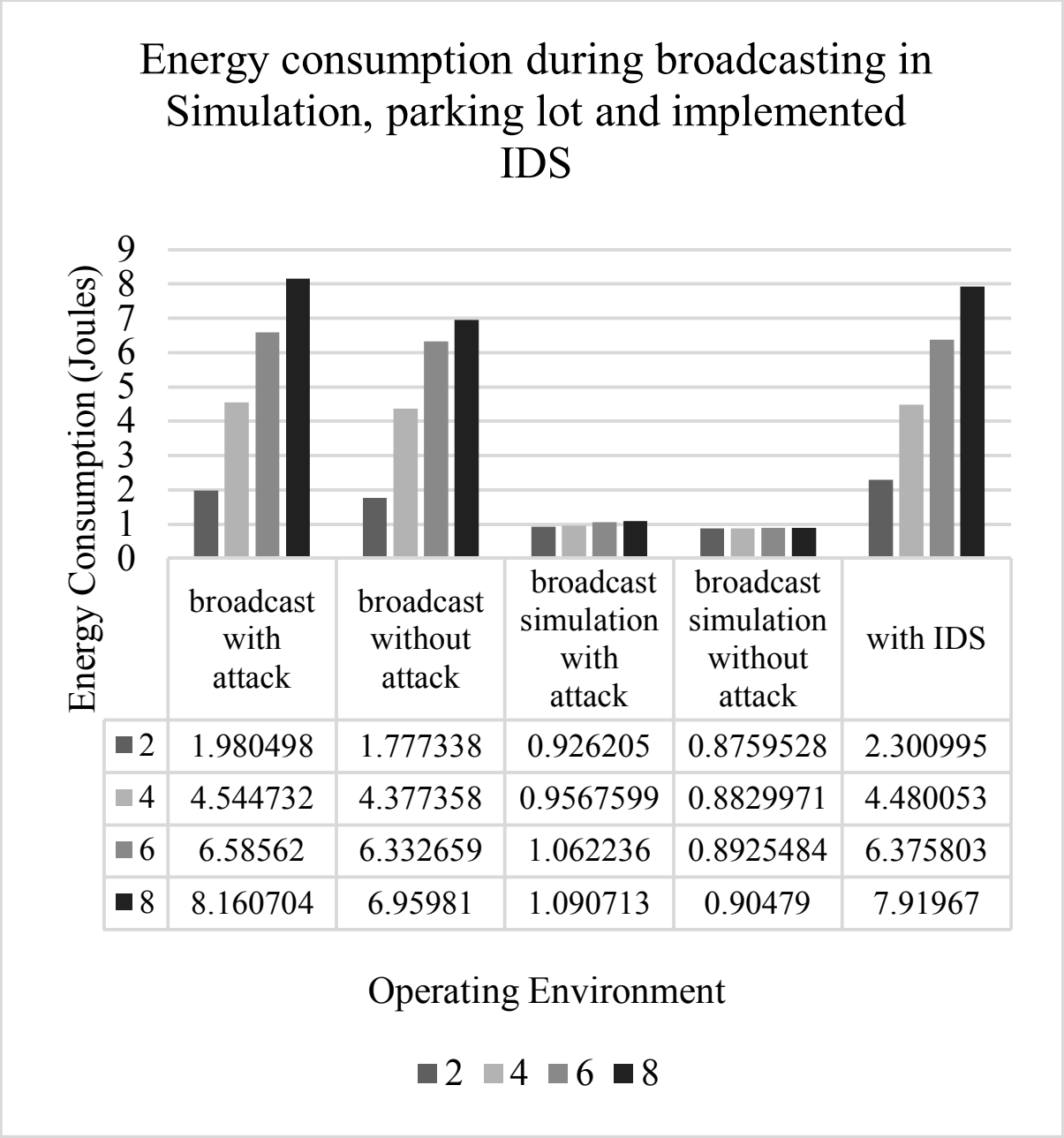

Figure 6. Energy Consumption During One to One communication in Simulation and lab space

ii. If corrupted node is 6BR then, monitoring of the whole IPv6 stack initiated

d. RSSI monitoring is initiated
e. Determine the RSSI values
f. Determining the nodes that are in the range considering the error while determining RSSI values.
g. Node which not in range while considering the errors in RSSI values considered as suspected node
h. Again, running the algorithm to check the probability of being that is suspected node
i. Higher the probability, higher the chances of being attacker node.
j. Then, suspected node is declared as Attacker node.

While detecting the attacker node, we can even track the node location as this algorithm is based on node localization. Following the algorithm implemented. As we can see that we have to follow every other step mentioned above.

The implemented algorithm is modified version of the existing IDS which was designed for wormhole attack considering the low powered devices or IoT devices such as Zolertia Z1. As we are analyzing the impact of implementation of intrusion detection system as these implementations should be efficient considering resource consumption. After the implementation, this algorithm is successful in detecting the cyber-attack, but we should also be concerned about the energy consumption of these devices and their battery life estimation.

Therefore in next section, we are focusing on the impact analysis of the proposed algorithm by conducting many experiments by varying the operating environment such as working lab space and open parking lot.

## IV. Results And Discussion

In this section, we are presenting the results of our impact analysis after the implementation of the algorithm.

As our most important aspect is to provide security for these devices but as we are working with such low powered devices, their impact analysis is also one of the most important things that we have to consider. We are conducting an experiment and presenting the results regarding the impact on these devices after the implementation of cyber-attack mitigation algorithm in terms of resource or energy consumption. The procured parameters from the powertrace example we run are in the form of CPU, LPM, Tx and Rx time.

Table 3. Operating States of Zolertia Z1 [13]

| Operating States | Ratings | Unit |
|---|---|---|
| **MCU on** | 18.8 | mA |
| **MCU on** | 17.4 | mA |
| **MCU idle** | 0.1 | µA |
| **MCU standby** | 0.5 | µA |
| **Voltage** | 3.6 | V |

Through powertrace code, we can evaluate the efficiency of the algorithm based on the accurate data regarding the power consumption of the whole network. To calculate the energy consumption in millijoules, we used the following standard equation [7] from literature, considering the parameter from Table 3.

$$Energy(mJ) = \frac{CPU * 0.5 + LPM * 0.0005 + Tx * 17.4 + Rx * 18.8) * 3}{32768}$$

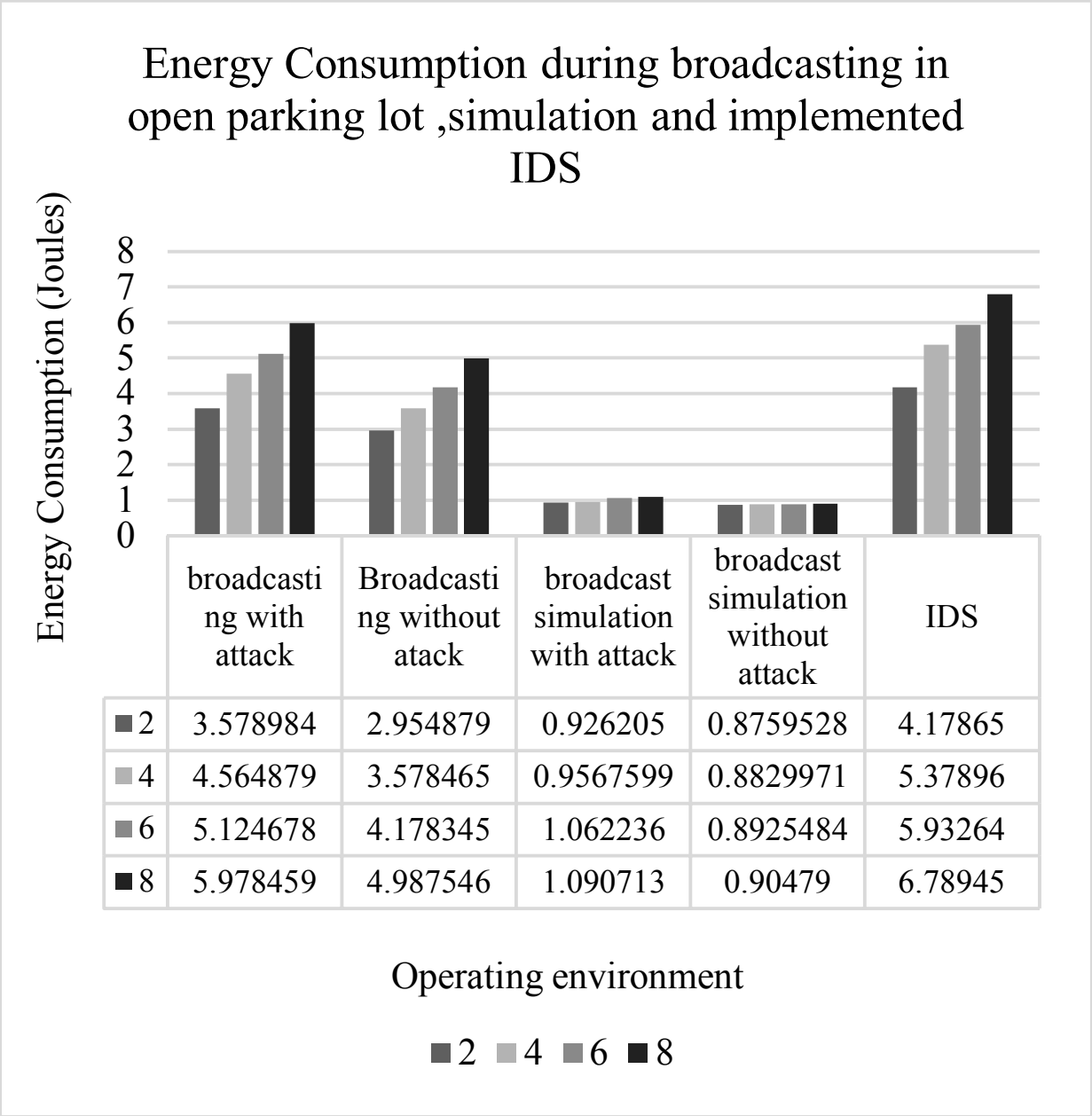

| | broadcasting with attack | Broadcasting without atack | broadcast simulation with attack | broadcast simulation without attack | IDS |
|---|---|---|---|---|---|
| 2 | 3.578984 | 2.954879 | 0.926205 | 0.8759528 | 4.17865 |
| 4 | 4.564879 | 3.578465 | 0.9567599 | 0.8829971 | 5.37896 |
| 6 | 5.124678 | 4.178345 | 1.062236 | 0.8925484 | 5.93264 |
| 8 | 5.978459 | 4.987546 | 1.090713 | 0.90479 | 6.78945 |

Figure 7. Energy Consumption during broadcasting in open parking lot and simulation

*Where*,
CPU = Time for which mote was active,
LPM = Total time for which the mote was in low power mode,
Tx = Total transmission time,
Rx = Total listening time.

In the topology we have used, the central node acts as an attacker node. The results of proposed algorithm were also compared with other existing algorithms to prove its power-efficiency. As an example, the power consumption analysis of our IDS indicates that although the real-world power consumption is quite high compared to simulation results, the algorithm results in lesser power consumption when compared with the system under attack. In figure 5, we can compare the energy consumption with and without attack, in simulation and on real world motes and with implemented IDS. We considered working lab space as an environment to conduct experiment primarily. We can clearly observe that Energy consumption increases significantly as the number of nodes is increasing. With IDS implemented there no significant increase in energy consumption.

In figure 6, we can again compare the energy consumption during one-to-one communication in simulation and lab space. The results are congruent to the one we have in figure 6. Interference present in the environment can also contribute to these changes in energy consumption. In both of these, we ran broadcasting example as we change the operating environment for both of that experiment. We can observe that from readings we accumulated in lab space and parking lot there is a noticeable difference in both those operating environment as there is much interference in parking lot compared to lab space. Also, we can notice that there is also a surge in resource consumption because of implementation along with interference present.

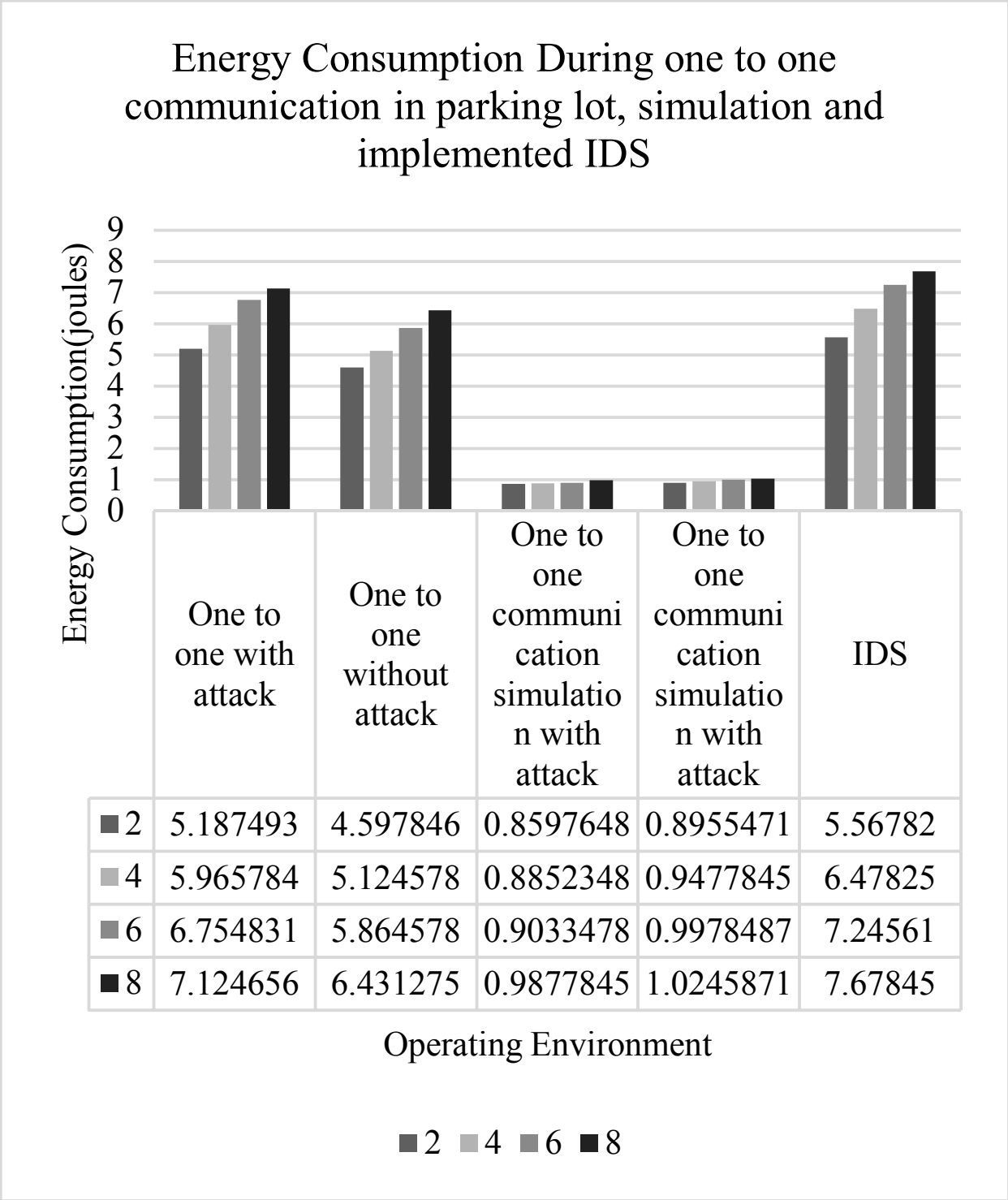

| | One to one with attack | One to one without attack | One to one communication simulation with attack | One to one communication simulation with attack | IDS |
|---|---|---|---|---|---|
| 2 | 5.187493 | 4.597846 | 0.8597648 | 0.8955471 | 5.56782 |
| 4 | 5.965784 | 5.124578 | 0.8852348 | 0.9477845 | 6.47825 |
| 6 | 6.754831 | 5.864578 | 0.9033478 | 0.9978487 | 7.24561 |
| 8 | 7.124656 | 6.431275 | 0.9877845 | 1.0245871 | 7.67845 |

Figure 8. Energy Consumption during one-to-one communication in parking lot and simulation

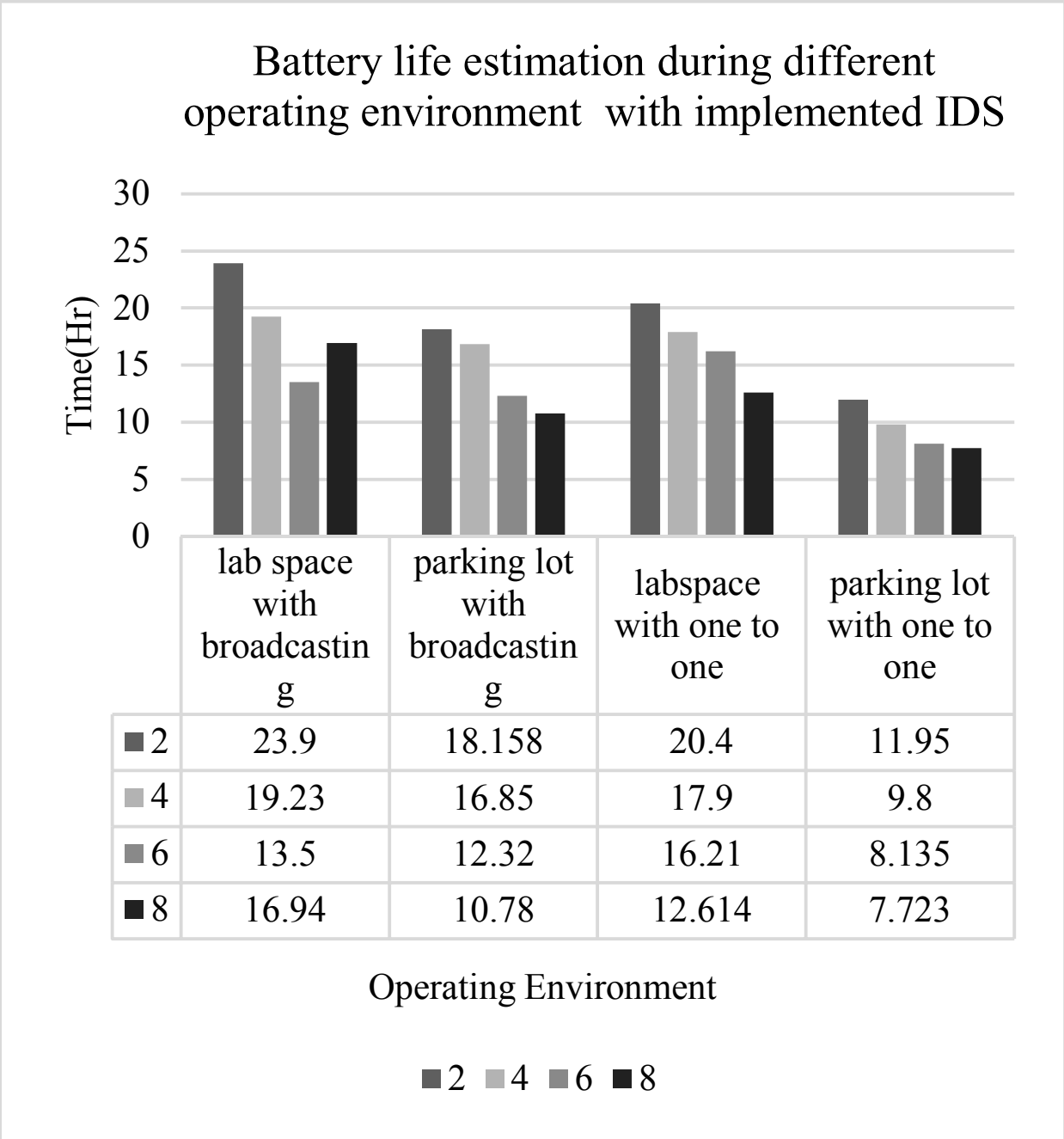

| | lab space with broadcasting | parking lot with broadcasting | labspace with one to one | parking lot with one to one |
|---|---|---|---|---|
| 2 | 23.9 | 18.158 | 20.4 | 11.95 |
| 4 | 19.23 | 16.85 | 17.9 | 9.8 |
| 6 | 13.5 | 12.32 | 16.21 | 8.135 |
| 8 | 16.94 | 10.78 | 12.614 | 7.723 |

Figure 9. Energy Consumption during different operating conditions

We can observe that from readings we accumulated in lab space and parking lot there is a noticeable difference in both those operating environment as there is much interference in parking lot compared to lab space. Also, we can notice that there is also a surge in resource consumption because of implementation along with interference present.

In figure 7 and 8, we can observe the results we accumulated from the conducted experiment considering open parking lot as an environment instead of working lab space. Similar to previous results the presented in these figures also consists of the effects of present interference. As for this experiment, we chose one to one communication example or Unicast example for communication. As for the one to one communication, we required lot of resource consumption contrary to our simulation results for the same example which we have noticed from our previously conducted experiments also [5]

As our previous study shows that interference present in the environment because of electrical interference, magnetic interference, etc., energy consumption increases significantly [5]. On the basis of the readings we got from the conducted experiments we can predict the battery life (T) of the devices by using the following equation:

$$E = \sum_{i=0}^{n} Vi * Ii * Ti$$

Where, E= energy in joules, I = current drawn, T = Time, V = Voltage, n is the number of overall readings.

With the help of this equation, we trace the battery consumption and can also predict the battery life. This equation [5] also give us detailed insight into the impact of the implementation on these devices.

In figure 9. We can get the exact idea about the battery life estimation of the nodes after the implementation of IDS on those nodes considering different environments such as working lab space and parking lot. As with the help of all the parameter measurements we got from all of the conducted experiments, we can estimate the battery life with the implementation of intrusion detection system. From the referenced figure, we can observe the estimation of battery life for lab space and open parking lot. In general, during broadcasting example with implemented IDS battery life is more compared to one to one communication. As from our conducted experiment we have concluded while IDS was already in place, consumption requirement for broadcasting in real world mote is less as compared to one to one communication. This can be clearly observed this significant difference in power consumption for these scenarios from figure 9.

## V. Conclusion

Considering the prevalence of these IoT devices, before we can employ these devices for a critical application, security aspects need to be addressed. More research in the area of IoT device security is needed to make IoT network efficient and secure. Therefore, an efficient and secure IDS was proposed in this paper for securing individual IoT devices. This algorith uses very low power, as seen from the results, and enhances security by detecting and thwarting cyber-attacks. The real-world experiments using Z1 Zolertia motes (first generation IoT devices) indicated the efficacy of proposed technique in different environments. As a limitation, it should be noticed that the algorithm might not work against all types of attacks. Therefore, generalization for thwarting more types of attacks could be a possible future work.